\documentclass[a4paper,usenatbib]{mnras}

\usepackage[T1]{fontenc}
\usepackage{ae,aecompl}

\usepackage{graphicx}	
\usepackage{amsmath}	
\usepackage{amssymb}
\usepackage{caption}
\makeatletter

\usepackage{latexsym}
\usepackage{setspace}
\usepackage{makeidx}
\DeclareGraphicsExtensions{.pdf, .png, .jpg}
\usepackage{graphicx, subfigure}
\usepackage[figurename=Figure]{caption}
\usepackage{multirow}

\title[]{The lensing time delay between gravitational and electromagnetic waves}

\author[P. Cremonese, E. M{\"o}rtsell]{
P. Cremonese$^{1}$\thanks{E-mail: cremonesep25@gmail.com} and 
E. M{\"o}rtsell,$^{2,3}$ 
\\
$^{1}$Dipartimento di Fisica e Astronomia Galileo Galilei, Padova\\
$^{2}$Department of Physics, Stockholm University\\
$^{3}$The Oskar Klein Centre, Stockholm University\\
}

\date{Accepted 2017. Received 2017; in original form 2017}

\pubyear{2018}

\begin{document}
\label{firstpage}
\pagerange{\pageref{firstpage}--\pageref{lastpage}}
\maketitle

\begin{abstract}
The recent detection of gravitational waves (GWs) and electromagnetic (EM) waves originating from the same source marks the start of a new multi-messenger era in astronomy.  
The arrival time difference between the GW and EM signal can be used to constrain differences in their propagation speed, and thus gravitational theories. We study to what extent a non-zero time delay can be explained by gravitational lensing when the line of sight to the source passes near a massive object. For galaxy scale lenses, this delay becomes relevant for GWs with frequencies between $10^{-6}$ and $10^{-9}$ Hz, sourced by super massive binary black-holes. In addition to GWs detectable by Pulsar Timing Arrays (PTAs), we expect to find also a unique and recognizable EM signal. 
We show that the delay between the GW and EM signal can be of the order of days to months; within reach of future observations. The effect may become important in future multi-messenger astronomy probing of gravitational propagation and interactions.  

\end{abstract}



\section{Introduction}

In September 2015, the first direct observation of GWs was made by the two detectors of the Laser Interferometer Gravitational-Wave Observatory (LIGO) \citep{Abbott:2016blz}. This observation opened a whole new field of study in astrophysics, and in the last years, research and interest in gravitational waves have increased considerably. 
As new observatories -- such as VIRGO (already in use) and Advanced VIRGO \citep{Banks:2017}, in Italy, the Japanese groundbased interferometer KAGRA (the KAmioka GRAvitational wave detector) \citep{Aasi:2013wya}, eLISA (the Evolved Laser Interferometer Space Antenna) \citep{Nishizawa:2016eza}, and DECIGO (the DECi-hertz Interferometer Gravitational wave Observatory) \citep{Isoyama:2018rjb, Yagi:2013du} -- are completed and begin operations, they will constitute new powerful instruments to study the universe.

With the first combined detection of gravitational and EM waves \citep{2017PhRvL.119p1101A}, a new multi-messenger era has begun in astronomy. 
In this work, we study the arrival time differences due to gravitational lensing, between GWs and EM signals emitted by a common source at the same time or with known intrinsic time delays. 
In \cite{Takahashi:2016jom}, two lens configurations -- point mass and singular isothermal sphere (SIS) -- were considered. It was found that the lens imprints a characteristic modulation on the waveform. In particular, if the wavelength of the GW is large compared to the gravitational radius of the lens, it will pass almost unperturbed, and there will be a time delay delay between the GW and EM signal.

A time delay between \textit{GW170817} and its EM counterpart is not expected for lenses with $M\gtrsim 10^3 M_\odot$ because of the large GW frequency. Note that if we consider modified gravity theories, GWs may propagate slower than the speed of light and there could be a time delay for this reason. The close to simultaneous detection of \textit{GW170817} and the EM signal has been used to constrain such theories \citep{Lombriser:2016,Creminelli:2017sry,Sakstein:2017xjx,Ezquiaga:2017ekz,Baker:2017hug,Boran:2017}.

A pulsar was used for the first indirect observation of GWs when in 1974, the energy loss of the binary system PSR 1913+16 was attributed to the emission of GWs \citep{1975ApJ...195L..51H}. Observations agreed with the theoretical expectation of general relativity to better than 0.1\,\%.

A different method involving pulsars being developed, and already in use for some time, to observe GWs with very low frequencies is Pulsar Timing Arrays (PTAs) \citep{1742-6596-610-1-012017, Lee:2011et, 2018MNRAS.477..964K, Huerta:2015, Sesana:2012}. 
The goal of this work is to compute and understand the feasibility of observing the time delay between EM signals and GWs observed with PTAs, corresponding to wavelengths longer or comparable to the gravitational radius of galaxy lenses. 
To have a time delay, there has to be a lens --  in our case typically a galaxy with $M\sim10^{11} M_\odot$ -- close enough to the line of sight to the source. The probability that a source at redshift $z\sim1$ is gravitationally lensed by a galaxy is of the order of $10^{-2}-10^{-3}$ \citep{Turner:1984}. 
In such a case, we find that the time delay will be observable for a large range of sources  for next generation observatories.

The article is organized as follows: In Section \ref{optics}, we explain why we expect a time delay and discuss the correct use of geometrical and wave optics. The heart of the work is Section \ref{time delay}, where we calculate and give numerical examples of the expected time delays . In Section \ref{Gravitational Wave Detection}, we describe PTAs and update on the latest results. Section \ref{SMBBH} treats the formation and evolution SMBBHs, and their GW and EM emission. The sensitivity needed to observe the time delays are presented in Section \ref{sensitivity} and conclusions are drawn in Section~  \ref{conclusions}. 

\section{Optical approximations}\label{optics}
For a gravitationally lensed source, a time delay between the gravitational and EM waves may result from their different wavelengths. Geometrical (or ray) optics and wave (or physics) optics are two approximations useful to study the propagation of waves. For a galaxy mass lens, GWs with frequencies in the range between $10^{-6}$ Hz and $10^{-9}$ Hz can be modelled using wave optics whereas EM waves can be described using geometrical optics. 
The large wavelengths of the GW will pass almost unperturbed through the lens and hence arrive first at the observer\footnote{Assuming that EM and gravitational signals leave the source at the same time.}. 

Before calculating the time delays, we define the boundary between the two approximations. 
\cite{2003ApJ...595.1039T} state that "in the gravitational lensing of gravitational waves, the wave optics should be used instead of the geometrical optics when the wavelength $\lambda$ of the gravitational waves is longer than the Schwarzschild radius of the lens mass" [see also \cite{1999PThPS.133..137N}]\footnote{A similar, but not equivalent, statement is given in \cite{gralen.boo}: "when the wavelength is larger than the path difference between the multiple images, the geometrical optics approximation breaks down". }. Therefore,
\begin{equation}
\begin{aligned}
\lambda\geq R_s&=\frac{2GM}{c^2}\approx\frac{GM}{c^2} \Rightarrow M\leq\frac{c^2}{G}\lambda= \\
&\simeq6,742\cdot10^{-4}M_{\odot}\left(\frac{\lambda}{\text{m}}\right)\approx10^5 M_{\odot}\left(\frac{f}{\text{Hz}}\right)^{-1} ,
\end{aligned}
\end{equation}  
and
\begin{equation}
M\leq10^5 M_{\odot}\left(\frac{f}{\text{Hz}}\right)^{-1}.
\end{equation}
Here, $R_s$ denotes the Schwarzschild radius of the lens.
For a galaxy with $M\sim10^{11}\ M_\odot$, if $\lambda\gtrsim 1.5\cdot10^{14}\ \rm{m}\Rightarrow\nu\lesssim 10^{-6}$ Hz, wave optics should be used. 
For $\nu\approx10^{-8}$ Hz, a typical GW frequency probed by Pulsar Timing Arrays (see chapter \ref{PTA}), wave optics must be used for lenses with $M<10^{13}\ M_\odot$. 

\section{Time delay}\label{time delay}
For a gravitationally lensed event where the wavelength of the GWs is longer than the Schwarzschild radius of the lens, the time difference is
\begin{equation}
\Delta T_{\rm EM,\pm -GW}(y,w)=T_{\rm EM,\pm}(y)-T_{\rm GW}(y,w),
\end{equation}
where $T_{\rm EM,\pm}$ and $T_{\rm GW}$ are dimensionless time delays of EM and GW signal, respectively. $y$ is the angular position of the source in units of the Einstein angle and $w$ is the dimensionless frequency of the GW (see \figurename~\ref{lens} and eqs. \ref{wdim}-\ref{timedeldim}). Details of the calculation are given in Appendix \ref{calculations}, and summarized in Tab. \ref{tdelaytab}. The arrival time difference is expressed in terms of the (dimensionless) phase difference between the two waves as $w\Delta T_{\rm EM,\pm -GW}$. We can reconstruct the dimensional time delay according to
\begin{equation}
\begin{split}
\Delta t_{\rm EM,\pm-GW}&=\frac{1}{2\pi f}w\Delta T_{\rm EM,\pm -GW}= \\
&=0.16 \cdot w\Delta T_{\rm EM,\pm -GW}\left(\frac{f}{\text{Hz}}\right)^{-1}\text{\ sec}.
\end{split}
\end{equation} 
For a \textit{point mass} lens, the maximum phase difference between the brighter EM image ($T_{\rm EM,+}$) and the GW for different source positions, $y$, are listed in Tab.~\ref{tmaxtable}. For $y=0.01$, $w\Delta T_{\rm EM,+ -GW}\simeq0.55$ and the maximum time delay is
\begin{equation}
\boxed{\Delta t^{\rm max}_{\rm EM,+ -GW}=0.09\text{\ sec}\left(\frac{f}{\text{Hz}}\right)^{-1}}\ .
\end{equation}
For $f\approx10^{-8}\text{ Hz (i.e.,} T\approx3$ yr), we have
\begin{equation}
\Delta t^{\rm max}_{\rm EM,+ -GW}\approx3.5 \text{ months.}
\end{equation} 
For $f\approx10^{-6}\text{ Hz (or } T\approx11$ days), $\Delta  t^{\rm max}_{\rm EM,+ -GW}\approx1$ day. \\
In the case of a SIS lens, for a source at $y=0.01$,
\begin{equation}
\boxed{\Delta t^{max}_{\rm EM,+ -GW}=0.1056\text{\ sec}\left(\frac{f}{\text{Hz}}\right)^{-1}}\ ,
\end{equation} 
slightly larger than for the case of a point mass lens. The large time delays emphasize the fact that, in these cases, gravitational lensing has to be considered. \\
Note that here, the dimensionless frequency $w$ is fixed. From eq.	\eqref{wdim}, $w\sim Mf$, and for a fixed $w$, the product of $f$ and $M$ is also fixed. In the point mass lens example, $\Delta t^{\rm max}_{\rm EM,+ -GW}\approx3.5$ months for a lens with\footnote{ From eq. \eqref{wdim}, $w=1.3\cdot f/\text{Hz}\cdot M/10^4 M_\odot$. Here, the lens redshift is $z=0.1$.} $M\approx8.3\cdot10^{11}\ M_\odot$.  
If the lens mass is constrained from other observations, we can constrain $w$. If not, we may use time delay observations to constrain its mass. 

\section{Gravitational Wave Detection}\label{Gravitational Wave Detection}
In this work, we study gravitational waves with typically $f\sim 10^{-8}$ Hz (that is $\lambda\sim 1$ pc and $T\sim1$ year), originating from super massive binary black holes (SMBBHs).

\subsection{Pulsar Timing Array}\label{PTA}
GWs with very low frequencies may be detected through the study of millisecond pulsar (MSPs), using Pulsar Timing Arrays (PTAs). The observed beam frequency of several MSPs are monitored over a long period of time and the pulse is modelled as precisely as possible and compared with the observed time of arrival (TOA) of the pulse. The difference is named the \textit{timing residual}. 
As in laser interferometry detectors, a GW passing between the pulsar and the Earth warps space-time, inducing a modulation in the TOA with respect to the model\footnote{ For further explanation see chapter \ref{emissionpropreities}.}. \\
The timing residual is usually different from zero because of noise. The noise is called \textit{red} or \textit{white}, depending on whether its power spectral density increases with frequency or stays flat, respectively. \textit{White noise} is due to radiometer noise, pulse jitter, and interstellar oscillation, while timing noise, dispersion measure (DM) variation, and an actual GW signal cause \textit{red noise}. Note that the gravitational signal is here classified as noise. Assuming we can correct for all unwanted sources of noise, what is left, in term of the timing residual, indicates a GW signal. 
With PTAs, it is possible to (i) increase signal-to-noise ratio of the GW in the timing residual \citep{DeCesar:2018}; (ii) compare the timing residual from different pulsars to discriminate between a GW signal and other noises \citep{1742-6596-610-1-012017}; (iii) reconstruct the position in the sky of the source \citep{sourceposition, 2018MNRAS.477.5447G}.

\subsubsection{Current and future results}\label{GWobservation}
Given the low frequencies involved, observations are time demanding and there are no GW signal currently detected using PTAs. Nonetheless, data collected so far are used to 
\begin{itemize}
\item improve noise detection and correction;
\item refine pulsar and GW source models; 
\item constrain the GW background (GWB).
\end{itemize}
Although the GWB is not yet detected, observational limits are getting close to the expected signal \citep{Lentati:2015qwp}. Since PTA instruments and data processing are improving quickly, there is optimism about making a detection in the next years \citep{DeCesar:2018}. A GWB detection would give information about the SMBBH population and other sources of GWs with frequencies of the order of nHz \footnote{The detection of single source GW signals will be discussed later.}.

A game changer in this field will be the Square Kilometre Array (SKA). When complete, the SKA \citep{Lazio:2013} will have a sensitivity corresponding to a telescope with a square kilometre mirror, and (i) a frequency range between 50 MHz (6 m) and 14 GHz (0.02 m), (ii) the highest sensitivity for a radio telescope (more than 50 times more sensitive than the best current telescopes) and (iii) a wide field of view \citep{Shao:2014wja}. \\
SKA has the potential to make huge contributions to many different aspects of radio astronomy. For what concerns this paper, it will increase the number of known pulsars (SKA2 potentially  could detect all galactic radio emitting pulsars in the SKA sky, beaming in our direction) and study them with unprecedented precision, as well as increasing data quality of the already known pulsars. It will also reduce the uncertainties in the position of the GW sources in the sky. This will allow us  to study GWs with a precision much higher than current PTAs \citep{Moore:2014lga}.   

\section{Super Massive Binary Black Holes}
\label{SMBBH}
Super massive black holes (SMBHs) are known to be present in the core of most massive galaxies \citep{MBHinGalaxies} and super massive binary black holes (SMBBHs), are believed to form in galaxy mergers. SMBHs in the core of the merging galaxies may create a binary system, stay in a prolonged in-spiral phase and eventually merge together. This process could last for millions of years\footnote{ A useful summary of the evolution of these kind of systems can be found in chapter 3 of \cite{Tanaka:2013o}}. 

Since the rate of these binary systems is not known, there is no consensus on whether we should expect single systems to emerge from the background radiation by being sufficiently close and/or very bright in GWs \citep{Sesana:2008xk}. In a recent paper \citep{2018MNRAS.477..964K}, it is argued that single sources are at least as detectable as the GW background.
   
\subsection{Characteristics of the emitted gravitational signal}
\label{emissionpropreities}

\begin{figure}
\centering
\includegraphics[width=0.49\textwidth]{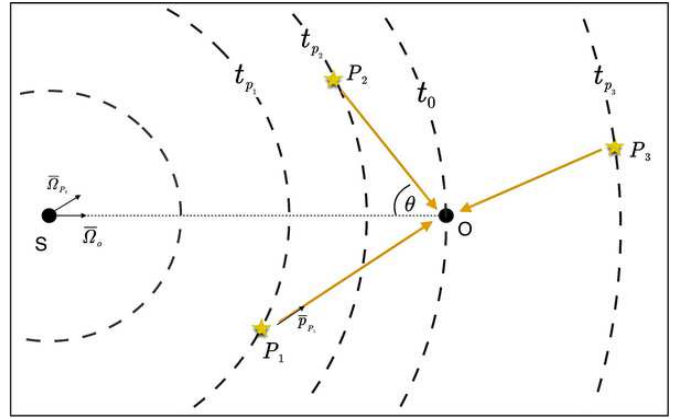}
\caption{Schematic example of the system composed by a GW source, S, a PTA with three pulsars, $P_1$, $P_2$ and $P_3$, and the Earth, O. Notice that the angle between $\bar{\Omega}_o$ and $\bar{\Omega}_P$ is exaggerated in the figure. Usually it is very small because the distance pulsar-Earth ($\sim 1-10$ kpc) is much smaller than the distance source-Earth ($\sim 10^2-10^3$ Mpc). For more details see the text.}
\label{PTAsystem}
\end{figure}

Taking into consideration the simplest binary configurations, the frequency emitted by SMBBHs is twice the orbital frequency. As the system loses energy, it enters an in-spiral phase where the main energy loss is due to GW emission. PTAs will be capable of detecting GWs from in-spiraling binaries with $m_{1,2}\geq10^8\ M_\odot$\footnote{ Further details in \cite{Sesana:2010ac}}, where $m_{1,2}$ are the SMBH masses. If we consider a pulsar emitting a radio pulse with frequency $\nu_0$, 
the space-time perturbation $h_{\alpha\beta}(t)$ induced by the GW, gives rise to an observed frequency shift. This is described by the characteristic two-pulse function
\begin{equation}
z(t,\bar{\Omega})\equiv\frac{\nu(t)-\nu_0}{\nu_0}=\frac{1}{2}\frac{\bar{p}^\alpha\bar{p}^\beta}{1+\bar{p}^\alpha\bar{\Omega}_\alpha}\Delta h_{\alpha\beta}(t,\bar{\Omega}),
\label{freqshift}
\end{equation}  
where $\nu(t)$ is the pulsar frequency received on Earth, $\bar{\Omega}$ is the unity vector parallel to the direction of the propagation of the GW, $\bar{p}$ is the unity vector indicating the propagation direction of radio waves from the pulsar and $\Delta h_{\alpha\beta}(t)\equiv h_{\alpha\beta}(t_p,\bar{\Omega})-h_{\alpha\beta}(t,\bar{\Omega})$ is the metric perturbation difference at the pulsar and at the observer, respectively (see \figurename\ \ref{PTAsystem}). The resulting time residual is given by
\begin{equation}
r(t)=\int_0^tdt'z(t',\bar{\Omega}).
\end{equation}
Since the frequency shift involves the metric perturbations both at the pulsar and the Earth, we expect the time residual for every pulsar to have two different terms. 
The Earth term, is due to GWs passing the Earth, and the pulsar term is due to GWs passing the pulsar. For a given source, the Earth term is the same for all the pulsars in the array. The pulsar term is different for each pulsar and depends also on the distance between the pulsar and the Earth, which, for the moment, is often poorly constrained. 

We expect the pulsar term to have different frequencies between each other and with the one coming from the Earth term. This is because GWs observed at the same time in the Earth and pulsar terms, have left the source at different times\footnote{ One can think of the pulsar term as -- for an EM signal -- having a mirror array in the sky which reflects the light coming from different sources in the Universe. These mirrors, being at distances of 1-10 kpc, would show us the sources at different ages.} (see \figurename\ \ref{PTAsystem}) and we expect the emitted signal to evolve with time. It can be shown \citep{Sesana:2010ac}, that the Earth and pulsar terms should be observable and distinguishable. The former will always be better determined since pulsars in the array can be used to improve the signal-to-noise, $S/N$, ratio. In fact, for this reason, the pulsar term is often ignored.   

\subsection{Electromagnetic signal}
\label{EMsignal} 
Articles summarizing the evolution of SMBBHs [\cite{Tanaka:2013o} and \cite{McKernan:2013} and citations therein], stress the fact that the models have large uncertainties. 
\begin{figure}
\centering
\includegraphics[width=0.48\textwidth]{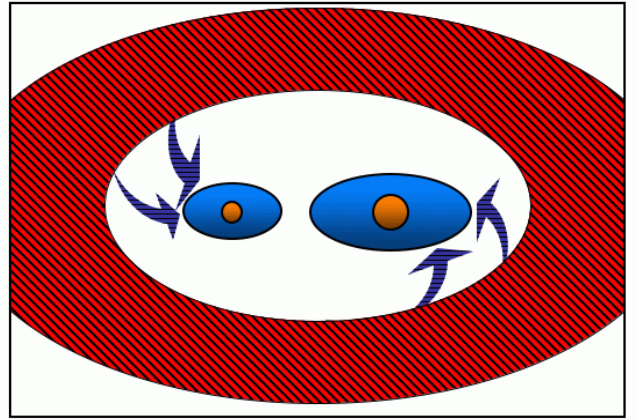}
\caption[]{Circumbinary disk scenario in which binary torques create a low density region in the center of the disk. The accretion onto the binary members is shown by the arrows. For more details see the text. Figure from \cite{Bogdanovic:2010}.}
\label{circumbinary}
\end{figure}
In order to create a SMBBH with GW emission observable via PTA, the SMBH masses have to be comparable, i.e. $0.01\lesssim q < 1$, where $q=m_2/m_1$ with $m_2\le m_1$. If $q<0.01$, the smaller galaxy may be tidally stripped and the SMBH inside it will not reach the center of the new forming galaxy\footnote{ Other articles, like \cite{Bogdanovic:2014}, set the lower limit to $q\approx0.1$.}. The secondary (i.e. the SMBH with lower mass) will create an annular gap around its orbital path\footnote{ This configuration is similar to the one of a hot Jupiter opening a gap in a protoplanetary disk.}. Eventually, the gas interior to the secondary's orbit will fall into the central SMBH and a cavity will form, as shown in \figurename\ \ref{circumbinary}. As gas continues to accrete, it leaks periodically into the cavity and may create a disk around one or both SMBHs. This is shown by the arrows and blue disks in \figurename\ \ref{circumbinary}. 

\begin{figure}
\centering
\includegraphics[width=0.47\textwidth]{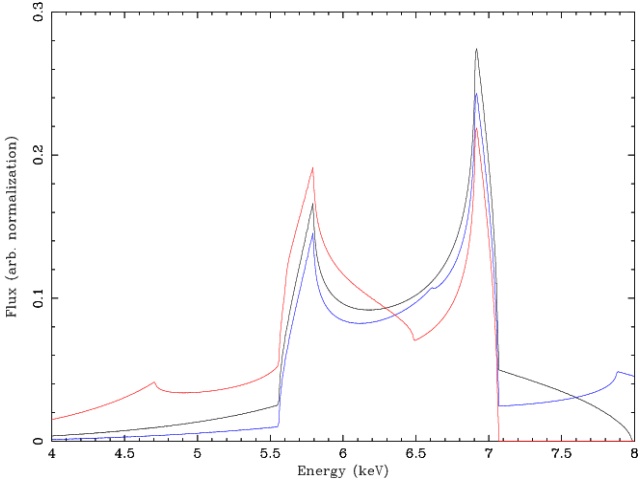}
\caption[]{The black curve shows the Fe K$\alpha$ emission line from the circumbinary disk (55 - 100 $r_g$) plus a weak secondary broad component (10\% of the intensity of the full disk profile) due to an accretion disk around the secondary black hole, located at 30 $R_g$, centered on the line centroid energy (6.40 keV). The red curve shows the effect of shifting the centroid of the weak secondary component redward to 5.2 keV. The blue curve shows the effect of shifting the centroid of the weak secondary component blue-ward to 7.3 keV. Figure from \cite{Tanaka:2013o}.}
\label{dopplerFe}
\end{figure}

It is debated how long these inner disks live. If they are stable and exist until the coalescence of the BHs [e.g., \cite{Kulkarni:2015}], this configuration is expected to have particular EM signals. The one we are interested in is due to Doppler effects on the emission line coming from the emitting disks, circumbinary and/or around one or both BHs. We consider the Fe K$\alpha$ line since it is one of the strongest and most studied. If there are no disks around the central BHs, the spectral profile of the line is constant with time, corresponding to the black line in \figurename\ \ref{dopplerFe}. If one of the BHs has an accretion disk, its Fe K$\alpha$ spectral profile will change with time \footnote{Unless the observer is observing the system face-on, in which case the velocity of the BHs along the line of sight is zero.}. The period of this change depends on the orbital period of the binary, while the line flux also depends on the mass ratio $q$, the distance between the BHs and the characteristics of the disk. In \figurename\ \ref{dopplerFe}, the line profile at maximum red-shift (red line) and blue-shift (blue line) are shown. The line is expected to change from one to another in half the period of the binary. From a prolonged observation, it could thus be possible to measure the binary period.

In \figurename~\ref{dopplerFe}, the mass ratio is $q\approx0.01$. If the two BHs are of comparable mass, as expected for binary systems emitting observable GWs, both BHs may have accretion disks. If that is the case, we expect the horns of the line to "pulse" over half the orbital period, since the time evolution of the accretion disks are similar but in opposition. 

\subsubsection{Detectability of EM counterparts}\label{detEMcount}
The detectability of the time-dependent line shape depends on many different characteristic of the system \citep{Sesana:2012}, but overall there are good chances that they are detectable. \cite{Tanaka:2013o} conclude that the spectral time variation would have been easily detectable during an extended observation with \textit{Astro-H}\footnote{ The paper was written before the unsuccessful launch of \textit{Astro-H}.}. \cite{McKernan:2013} argues that also \textsc{EPIC pn} onboard \textit{XMM-Newton}, through repeated observations should be able to detect the line oscillations. Both works refer to an AGN at redshift $z=0.01$.
The SKA will allow us to resolve individual SMBBHs emitting GWs in the PTA range \citep{Sesana:2012}. The number and location of these sources depends on MBHs formation models. \\
There are already examples of observations, suggesting the presence of binary systems similar to those just portrayed\footnote{ See, for example, \cite{Graham:2015}, \cite{Bon:2016}, \cite{Liu:2016}, \cite{Kun:2014}}. However, none of them provide conclusive evidence for SMBBHs.

\section{Sensitivity of observations}
\label{sensitivity}
We now estimate the $S/N$ ratio needed to detect GWs from a single source, as well as measure the time delay. 
\cite{Cutler:1994} and \cite{Takahashi:2016jom} show that, in a matched filtering analysis, the phase of the waveform can be measured with an accuracy corresponding to the inverse signal-to-noise ratio, $\approx(S/N)^{-1}$. For example, if $S/N=10$, we can measure the phase difference if $\omega\Delta T_{\rm EM,\pm -GW}\gtrsim 10^{-1}$ rad. Conversely, to detect a phase difference of $\approx0.11$, corresponding to the lowest value in Tab. \ref{tmaxtable},  we need $S/N\gtrsim(0.11)^{-1}\simeq9.1$. Even though both GW and EM phases enters in $w\Delta T^{max}$, GWs will dominate the uncertainty, and we concentrate on determining the $S/N$ for GWs from PTAs detections. \\
Following \cite{Huerta:2015}, in Appendix \ref{snr} we derive the $S/N$ ratio  
\begin{equation}
\rho_{\rm high}^2=\hat{\rho}^2\cdot (1+z)^4\left(\frac{f_{\rm orb}}{f_{\rm obs}}\right)^{-2/3},
\label{snreq}
\end{equation}
where $f_{\rm orb}$ is the orbital frequency of the SMBBH, $f_{\rm obs}$ is the lowest frequency detectable by the PTA, and $\hat{\rho}^2$ (defined in eq. \ref{rhohat}) depends on the mass of the binary system, the total baseline time of observation, the distance of the source, the root mean square of the timing noise and on the cadence of observations.

In the next examples, and in Tab. \ref{SNRtab}, we consider sources at redshift $z=1$, a luminosity distance of $d_{\rm L}\simeq6.7$ Gpc, a total observing time of $T_{\rm obs}=10$ yr (and therefore $f_{\rm obs}=2\cdot3.17\cdot10^{-9}\text{ s}^{-1}$), an observing cadence of $1$ week, i.e. $\Delta t\simeq0.02$ yr, and a timing noise with $\sigma\approx100$ ns.

For the international PTA (IPTA)\footnote{ The IPTA is the collaboration between the three major PTA collaborations \citep{2016IPTA}} with $N_p\simeq30$ pulsars, and a SMBBH with $m_1=m_2=10^8\ M_\odot$, we obtain a signal-to-noise ratio $\rho^2\approx0.24$ for $f=2f_{\rm orb}=10^{-8}$ Hz. For $f=2f_{\rm orb}=10^{-6}$ Hz,  $\rho^2\approx0.01$, too low to detect the largest time delay calculated in section \ref{time delay} (see Tab. \ref{tmaxtable}), or even to recognize the GW signal. 
Setting the minimum signal-to-noise ratio at $S/N\gtrsim5$, $m_1=m_2\gtrsim 6.3\cdot10^8 M_\odot$ for the time delay to be detectable for GW frequencies $f\lesssim10^{-6}$ Hz. 

With SKA, a larger numbers of pulsars will be detected and studied, increasing also the detectability of time delays. All other parameter values maintained, for an array with $N_p=500$ pulsars and $m_1=m_2=10^8 M_\odot$, we obtain $\rho^2\approx3.2$ for $f=10^{-6}$. Slightly increasing the SMBBH mass will make the observation feasible. Numerical examples are summarised in Tab. \ref{SNRtab}.
\begin{table}
\centering
\begin{tabular}{|c|c|c|c|}
\hline
\textbf{Frequency} (s$^{-1}$) & $\boldsymbol{N_{\rm p}}$ & \textbf{Mass} ($M_\odot$) & $\boldsymbol{S/N}$ \\
\hline
\multirow{6}*{$10^{-8}$} & \multirow{3}*{$30$} & $1.0\cdot10^8$ & 0.24 \\
 & & $2.5\cdot10^8$ & 5.3 \\
 & & $3.0\cdot10^8$ & 9.3 \\
\cline{2-4}
 & \multirow{3}*{$500$} & $3.5\cdot 10^7$ & 1.9 \\
 & & $5.0\cdot10^7$ & 6.6 \\
 & & $1.0\cdot10^8$ & 70 \\
\hline
\multirow{6}*{$10^{-6}$} & \multirow{3}*{$30$} & $1.0\cdot10^{8}$ & 0.01 \\
 & & $6.3\cdot10^8$ & 5.2 \\
 & & $8.0\cdot10^8$ & 11 \\
 \cline{2-4}
 & \multirow{3}*{$500$} & $5.0\cdot10^7$ & 0.3 \\
 & & $1.0\cdot10^8$ & 3.2 \\
 & & $1.5\cdot10^8$ & 12 \\
\hline
\end{tabular}
\caption{Signal-to-noise ratios as a function of SMBBH masses and the number of pulsars in the PTA, for the two limiting frequencies. Note that from eq. \eqref{snreq}, the $S/N$ is higher for lower frequency, all other parameters fixed.}
\label{SNRtab}
\end{table}

\section{Conclusions}
\label{conclusions}
For a gravitationally lensed source, there will be a time delay between GWs and EM signals, if the GW wavelength is much larger than the gravitational radius of the lens, which in turn is much larger than the wavelength of the EM signal.
This effect may become an important tool for future multi-messenger astronomy, e.g., to study the propagation speed of gravitational theories \citep{Fan:2017}, and thus gravitational theories. 

For the first combined observation of GWs with an EM counterpart, \textit{GW170817}, no delay is expected for lenses with $M\gtrsim 10^3 M_\odot$. 
In this paper, we note that for GWs signals detected with PTAs, a time delay to a possible EM counterpart is expected in cases where the source is lensed by galaxy mass objects.
In fact, both GW and EM signals from inspiraling SMBBHs are, in theory, detectable with current technologies, although not yet realized.
With future data from SKA, we will be able to measure the gravitational signal, coming from SMBBHs with total mass $M\gtrsim10^8\ M_\odot$, with a $S/N\gtrsim 3$. Together with EM observations from next generation $X$-ray satellites, whose sensitivity needs to be further studied for distant sources, this will allow us to measure time delays of the order of months, expected for a lens mass $M\approx10^{11}\ M_\odot$. 
Note that the lens mass can be constrained from the EM image positions.

A major challenge of these observations is the detection of GWs with high enough sensitivity. We showed that, with a large enough number of pulsars, and with prolonged and precise observations, this will be possible. For close sources, the EM counterpart should be easily detectable, while for sources at redshift $z\approx1$, it is not still clear how well this can be pursued (see chapter \ref{detEMcount}). The fact that the expected EM signals are unique for binary SMBHs systems, opens up for the possibility to recognize and identify these sources, and perform dedicated follow up GW observations. 

Once the theoretical feasibility of the observations are certified, an open problem is how accurate we can couple the emission of the GW and EM signals in the time domain. We expect a direct correlation between the frequency of the GW and the frequency of EM spectral line modulation. The study of whether observations will allow this to be done with sufficient precision is left for future work. 


\nocite{*}
\bibliographystyle{mnras}
\bibliography{biblio} 


\appendix

\section{Time delay calculations}
\label{calculations}
In geometrical optics, the time delay function is 
\begin{equation}
t(\vec{\theta},\vec{\beta})=\frac{1+z_{\rm d}}{c}\frac{D_{\rm d}D_{\rm s}}{D_{\rm ds}}\left[\frac{1}{2}(\vec{\theta}-\vec{\beta})^2-\psi(\vec{\theta})\right],
\label{timedel}
\end{equation}
where $\theta$, $\beta$, $D_d$, $D_s$ and $D_{ds}$ are defined in \figurename~\ref{lens}, and $\psi(\vec{\theta})$ is the effective lensing potential \citep{gralen.boo}. For GWs, we need to use wave optics, and the calculations are more involved. We will derive here the time delay for point masses and SIS lenses.

\subsection{Lensed GWs} 
\label{lensedGW} 
\begin{figure}
\centering
\includegraphics[scale=0.5]{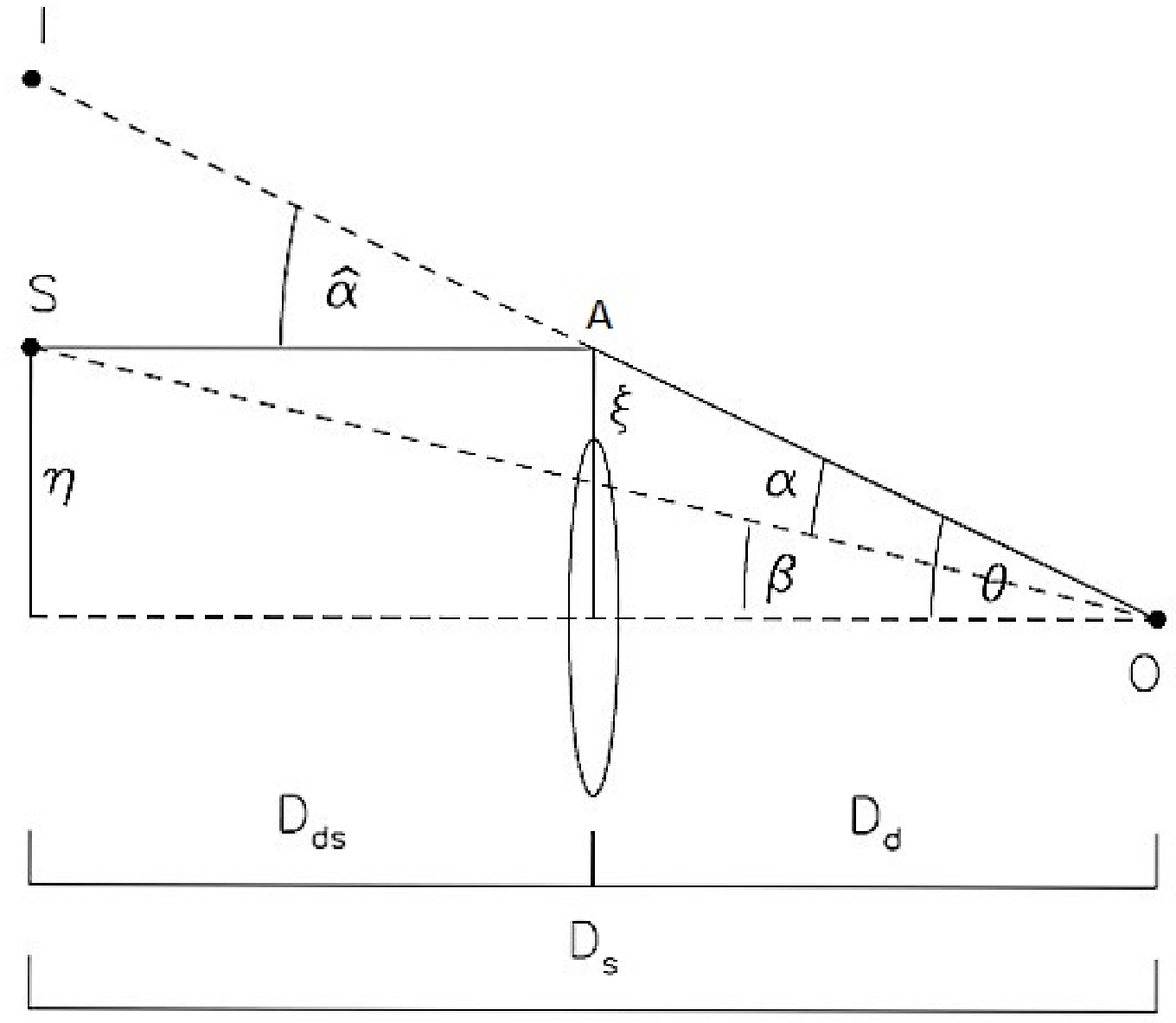}
\caption[]{Geometry of a gravitational lens system. S is the source and O the observer. The angular diameter distances between observer and lens, lens and source, and observer and source are $D_{d}$, $D_{ds}$, and $D_{s}$, respectively. $I$ is the position of the source as seen from the observer. Figure from \cite{gralen.art}.}
\label{lens}
\end{figure}

To calculate the lensed form of GWs, $\tilde{h}^L_{+,\times}(f)$, we use the \textit{amplification factor}. This is a complex function, $F(f,\beta)$, given by the diffraction integral [derived in \cite{gralen.boo}],
\begin{equation}
F(f,\beta)=\frac{D_{\rm d}D_{\rm s}}{cD_{\rm ds}}\frac{(1+z_{\rm d})f}{i}\int d^2\theta \exp[2\pi ift_{\rm d}(\theta,\beta)],
\end{equation}
where $f$ is the frequency of the GW\footnote{ In this section, we consider monochromatic GWs.}, $\beta$ is defined in \figurename~\ref{lens} and the time delay, $t_{\rm d}(\theta,\beta)$, is defined in eq. \eqref{timedel}. 
The lensed GW is given by the product of the unlensed waveform, $\bar{h}_{+,\times}(f)$, and the amplification factor 
\begin{equation}
\tilde{h}^L_{+,\times}(f)=F(f,\beta)\cdot\bar{h}_{+,\times}(f).
\end{equation}
Defining $x\equiv\theta /\theta_{\rm E}$ and $y\equiv\beta /\theta_{\rm E}$, we can define 
\begin{subequations}
\begin{align}
w&\equiv\frac{D_{\rm d}D_{\rm s}}{cD_{\rm ds}}\theta_{\rm E}^2(1+z_{\rm d})2\pi f=8\pi\frac{GM}{c^3}f, \label{wdim}\\
\begin{split}
T(x,y)&\equiv\frac{cD_{\rm ds}}{D_{\rm d}D_{\rm s}}\theta_E^{-2}(1+z_{\rm d})^{-1}t_{\rm d}(\theta,\beta)= \\ &\qquad\qquad\qquad=\frac{c^3}{4GM}t_{\rm d}(\theta,\beta),\end{split} \label{timedeldim} \\
\notag
\end{align}
\end{subequations}
where the Einstein angle, $\theta_{\rm E}$, is given in terms of the lens mass, $M$, as 
\begin{equation}
\theta_{\rm E}^2=\frac{4GM_{\rm z}}{c^2}\frac{D_{\rm ds}}{D_{\rm d}D_{\rm s}}.
\end{equation}
The amplification factor can now be written
\begin{equation}
F(w,y)=\frac{w}{2\pi i}\int d^2x\exp[iwT(x,y)].
\label{amplificationfactor}
\end{equation}
The time delay of the GW is defined from the phase of the amplification factor,
\begin{equation}
T_{GW}(w,y)\equiv -\frac{i}{w}\ln{\left(\frac{F(w,y)}{|F(w,y)|}\right)}.
\label{TGW}
\end{equation}
Note that this delay, unlike the one for light, depends on the frequency of the wave. \\
For a point mass lens, the amplification factor is \citep{Takahashi:2003},
\begin{equation}
\begin{split}
F^{\rm p}(w,y)=\exp&\left[\frac{\pi w}{4}+\frac{iw}{2}\ln\left(\frac{w}{2}\right)\right]\Gamma\left(1-\frac{iw}{2}\right)\cdot \\ &\cdot\ _1F_1\left(\frac{iw}{2},1;\frac{iwy^2}{2}\right),
\end{split}
\end{equation}
where $_1F_1$ is the confluent hypergeometric function \citep{Peters:1974}.
For $w\ll1$, the GW time delay for a point mass lens is
\begin{equation}
\boxed{T^{\rm p}_{\rm GW}(w,y)=\frac{1}{2}\left[\ln\left(\frac{w}{2}\right)+\gamma\right]+\mathcal{O}(w^2)}\ ,
\end{equation}
where $\gamma=0.577215...$ is the Euler constant. \\
For a SIS lens, the time delay is
\begin{equation}
\boxed{T^{\rm SIS}_{\rm GW}(w,y)=-\frac{\sqrt{\pi}}{2}w^{-1/2}-\left(1-\frac{\pi}{4}\right)+\mathcal{O}(w^{1/2})}\ .
\end{equation}

\subsection{Lensed EM light} 
The EM time delay is calculated in the geometrical optics regime using eq. \eqref{timedel}, or in dimensionless form, using eq. \eqref{timedeldim}. \\
For a \textit{point mass} lens [see \figurename~\ref{lens}]
\begin{equation}
\vec{\beta}=\vec{\theta}-\vec{\alpha}(\vec{\theta})=\theta-\frac{\theta_E^2}{\theta},
\label{beta}
\end{equation}
and
\begin{equation}
\boxed{T_{\rm EM,\pm}(y)=\frac{y^2+2\mp y\sqrt{y^2+4}}{4}-\ln\left|\frac{y \pm\sqrt{y^2+4}}{2}\right|}\ .
\end{equation}
For an SIS lens, $\theta_\pm=\beta\pm\theta_E$, and
\begin{equation}
\boxed{T_{\rm EM,\pm}(y)=\mp y-\frac{1}{2}}\ .
\end{equation} 
Note that time delays can be negative. Usually, the time delay is defined to be zero for the unlensed case, greater than zero when the lensed signal arrives after the unlensed one and vice versa. However, this normalization is only possible if the lensing potentials can be normalized to be zero at infinite distance from the lens. This is not the case for the potentials employed in this paper. Specifically, the mass of a singular isothermal sphere diverges at infinite radius. 
Since we are not measuring the time delay with respect to the unlesed case, but only considering the delay between GW and EM signals, the normalization becomes unimportant and negative time delays are not a problem.

\begin{table*}
\centering
\begin{tabular}{|c|c|c|c|}
\hline
\textbf{Lens} & \textbf{GW delay} & \textbf{EM delay} & $\boldsymbol{\Delta t_{\rm EM,\pm-GW}}$\\
\hline
\multirow{2}*{Point mass} & $\frac{1}{2}\left[\ln\left(w/2\right)+\gamma\right]+$ & $(y^2+2\mp y\sqrt{y^2+4})/4+$ & $0.16$ sec $(f/\text{Hz})^{-1}\cdot$ \\
 & $+\mathcal{O}(w^2)$ & $-\ln\left|(y \pm\sqrt{y^2+4})/2\right|$ & $\cdot w\Delta T_{EM,\pm-GW}$  \\ 
\hline
\multirow{2}*{SIS} & $-(w^{-1/2}\sqrt{\pi})/2-\left(1-\pi /4\right)+$ & \multirow{2}*{$\mp y-1/2$} & $0.16$ sec $(f/\text{Hz})^{-1}\cdot$ \\
 & $+\mathcal{O}(w^{1/2})$ & & $\cdot w\Delta T_{EM,\pm-GW}$  \\
\hline
\end{tabular}
\caption{Time delays for point mass and SIS lens, depending on the source position, $y$, and (for GWs) on the wave frequency, $w$. The values of $w\Delta T^{max}$ for different $y$ are listened in Tab. \ref{tmaxtable}.}
\label{tdelaytab}
\end{table*} 

\begin{table*}
\centering
\begin{tabular}{|c|l|c|c|c|}
\hline
\textbf{Lens} & \textbf{y} & $\boldsymbol{w}$ & $\boldsymbol{w\Delta T^{max}}$ (rad) & $\boldsymbol{\Delta T^{max}\cdot(f/\text{Hz})^{-1}}$ (sec) \\
\hline
\multirow{3}*{Point mass} & $1$ & $0.23$ & $0.11$ & 0.018 \\
 & $0.1$ & $0.92$ & $0.46$ & 0.073 \\
 & $0.01$ & $1.10$ & $0.55$ & 0.088 \\
\hline
\multirow{3}*{SIS} & $1$ & $0.12$ & $0.15$ & 0.024 \\
 & $0.1$ & $1.32$ & $0.51$ & 0.082 \\
 & $0.01$ & $2.25$ & $0.66$ & 0.106 \\
\hline
\end{tabular}
\caption{Maximum time delays for different source positions, $y$, for point mass and SIS lenses. The $w$ are obtained setting to zero the derivative of $\Delta t_{\rm EM,+ -GW}(w,y)$ with respect to $w$, and the $\Delta T^{max}$ are obtained by inserting this value of $w$ in the relevant equations.}
\label{tmaxtable}
\end{table*}

\section{Signal to noise ratio calculations}
\label{snr}
\cite{Huerta:2015} consider binary systems with eccentricity $e\neq0$. In the limit of $e=0$, we have 
\begin{subequations}
\begin{align}
\rho_{\rm high}^2=&\ \hat{\mathcal{B}}\cdot f_{\rm orb}^{-2/3}, \text{ for } f\gtrsim\frac{2}{T_{\rm obs}}, \label{SN}\\
\rho_{\rm low}^2=&\ \hat{\mathcal{C}}\cdot f_{\rm orb}^{16/3}, \text{ for } f\lesssim\frac{2}{T_{\rm obs}},
\end{align}
\end{subequations}
where $T_{\rm obs}$ is the total baseline time of observation, and $f_{\rm obs}\sim T_{\rm obs}$ is the lowest frequency detectable by the PTAs\footnote{ Nonetheless, the case for $f\lesssim 2/T_{\rm obs}$ is usually taken into account for completeness, since SMBBHs with eccentricity $e\neq0$ emit a spectrum of different GWs wavelengths.}. $f_{\rm orb}$ is the orbital frequency of the SMBBH, and $\mathcal{B}$ and $\mathcal{C}$ are
\begin{subequations}
\begin{align}
\hat{\mathcal{B}}=&\frac{4\sqrt[3]{2}\pi^{4/3}N_{\rm p}(N_{\rm p}-1)}{45}\frac{T_{\rm obs}\mathcal{M}^{10/3}(1+z)^{4}}{d_{\rm L}^2\Delta t\sigma_{\rm rms}} ,\label{SNfactor}\\
\hat{\mathcal{C}}=&\frac{4\sqrt[3]{2}\pi^{4/3}N_{\rm p}(N_{\rm p}-1)}{45}\frac{T_{\rm obs}^7\mathcal{M}^{10/3}}{d_{\rm L}^2(1+z)^2\Delta t\sigma_{\rm rms}}.
\end{align}
\end{subequations} 
Here, $N_{\rm p}$ is the number of pulsars in the PTA, $d_{\rm L}$ the luminosity distance to the source, $\sigma_{\rm rms}$ the root mean square of the timing noise, $\mathcal{M}$ is the chirp mass\footnote{ The chirp mass of a compact binary system determines the leading order amplitude and frequency evolution of the emitted GWs.} of the SMBBH system, and $1/\Delta t$ is the cadence of the observations. For simplification, we rewrite eqs. \eqref{SN} and \eqref{SNfactor} as\footnote{ We just consider the case of $f\gtrsim2/T_{\rm obs}$.}
\begin{equation}
\rho_{\rm high}^2=\hat{\rho}^2\cdot (1+z)^4\left(\frac{f_{\rm orb}}{f_{\rm obs}}\right)^{-2/3},
\label{snreq}
\end{equation}
with
\begin{equation}
\begin{aligned}
\hat{\rho}^2=4.26\cdot10^{-2}N_{\rm p}(N_{\rm p}-1)\left(\frac{\mathcal{M}}{10^8 M_\odot}\right)^{10/3} \times \\ 
\times \left(\frac{T_{\rm obs}}{10\text{ yr}}\right)^{5/3}\left(\frac{100\text{ Mpc}}{d_{\rm L}}\right)^2\left(\frac{100\text{ ns}}{\sigma_{\rm rms}}\right)\left(\frac{0.05\text{ yr}}{\Delta t}\right).
\end{aligned}
\label{rhohat}
\end{equation}


\bsp	
\label{lastpage}
\end{document}